\documentstyle[12pt]{article}
\date{}
\begin{document}
\title{Repulsive Short Range Three-Nucleon Interaction}
\author{S.\ A.\ Coon$^1$, M.\ T.\  Pe\~{n}a$^2$, and D.\ O.\  Riska$^3$}
\maketitle
\centerline{$^1$ \it Physics Department, New Mexico State University,
Las Cruces, New Mexico 88033, USA}
\centerline{$^2$ \it Centro de Fisica Nuclear, 1699 Lisboa and
Instituto Superior T\'ecnico, 1096 Lisboa, Portugal}
\centerline{$^3$ \it Department of Physics, University of Helsinki,
00014 Finland}

\setcounter{page}{0}
\vspace{1cm}

\centerline{\bf Abstract}
\vspace{0.5cm}
The three nucleon interaction that arises from pion and the effective
scalar and vector meson exchange components of the nucleon-nucleon
interaction is constructed and shown to be repulsive. Using several
wavefunction models we show that this interaction reduces the
calculated binding energy of the trinucleons by about 200 keV.
The contributions of intermediate $N$(1440) resonances
to this three nucleon interaction and of the $D$-state component
in the trinucleon model are estimated, and shown to be small.\\
\vspace{1cm}
Submitted to The Physical Review C\\
Research Institute for Theoretical Physics\\
University of Helsinki Preprint HU-TFT-95-11

\newpage
\centerline{\bf 1. Introduction}
\vspace{0.5cm}

Although the nuclear three-nucleon interaction (TNI) is very weak in
comparison to the two-nucleon interaction \cite{BG}, it nevertheless has
been found that the binding energies of the bound three- and
four-nucleon systems cannot be understood without taking into account
the attraction caused by the TNI \cite{reviews}. The main component of the
nuclear TNI is that associated with two-pion exchange, which arises
from,
for the major part, pion rescattering through an
intermediate virtual $\Delta_{33}$ resonance. The standard model for the
two-meson exchange component of the TNI is the so-called Tucson-Melbourne
model,
which also includes $\rho$-meson exchange in addition to pion exchange
\cite{TM1,TM2}. It has been found that when realistic models for the
nucleon-nucleon interaction are employed, the Tucson-Melbourne $\pi$-exchange
model for the TNI leads to an overbinding of a few
hundred keV in the trinucleons \cite{Chen} and of 2-4 MeV in
the case of the alpha particle \cite{overbind}. To compensate for this
overbinding an additional repulsive spin-independent phenomenological
TNI of short range has been proposed \cite{phTNI}. On the other hand,
the Tucson-Melbourne model ({\em with both $\pi$- and $\rho$-exchange}) has
recently been shown to give the correct binding of the tri-nucleon
with the Tucson-Melbourne meson-baryon-baryon vertex functions (``form
factors") \cite{Stadler}.  These form factors, however, differ in their
short range behavior from those of the (older) realistic models of the
nucleon-nucleon interaction. While
this form factor discrepancy \cite{GT} is further
studied in the case of the NN interaction \cite{Holinde}, there
remains a need for further investigation of the short range aspects of
the TNI.\\

There is no known dynamical mechanism that would lead to a
spin-independent short range TNI. The short range three-nucleon
interaction that arises from scalar and vector meson exchanges with an
intermediate nucleon-antinucleon pair were considered in
ref.\cite{Keister}, but were
found to be both insignificantly small and spin
dependent. We here consider another set of related three-nucleon
interactions - those that arise from pion and ``effective" scalar and
vector meson exchanges and that involve excitation of intermediate
nucleon-antinucleon pairs and $N(1440)$ resonances (Fig. 1). The
presence and form of the former (Fig. 1a) - and more important one of these
interactions - is implied by the pion and scalar and vector meson
exchange components of the nucleon-nucleon interaction, and can be
derived directly from a given complete model for the two-nucleon
interaction. The derivation is especially straightforward if the
nucleon-nucleon interaction has explicit scalar and vector meson
exchanges with associated meson-baryon-baryon vertex functions, but the
former TNI can also be constructed from ``effective" scalar and
vector meson exchanges obtained from any complete energy-independent
model of the nucleon-nucleon interaction \cite{Blunden}.
The magnitude of the latter TNI (Fig. 1b) is less certain because of
the wide uncertainty in the meson-nucleon-$N(1440)$ (``Roper resonance")
coupling strengths. Both of these sets of three-nucleon interactions
nevertheless have a very simple form, although spin dependent, and
when combined provide an amount of additional repulsion that
approximately corresponds to that required and hitherto ascribed to the
purely phenomenological spin-independent short range TNI. \\

In section 2 of this paper we derive the effective $\pi$-scalar and
$\pi$-vector meson exchange three nucleon interactions which arise
from excitation of intermediate nucleon-antinucleon pairs and show how
the potentials which describe these interactions can be constructed
from realistic models for the nucleon-nucleon interaction. It is this
construction method which distinguishes the TNI's of the present paper from the
Brown-Green \cite{BG}, Tucson-Melbourne \cite{TM1,TM2}, Brazil
\cite{Robilotta}, and the Fujita-Miyazawa \cite{FM} TNI's (the last being
extensively used in a program of studies with light nuclear systems
\cite{Argonne})  The latter TNI's did not attempt such a tight connection
with a nucleon-nucleon interaction but relied on other aspects of
hadronic phenomenology for motivation and parameter fixing.
In section 3
we derive the corresponding three nucleon interactions that arise from
excitation of $N(1440)$ resonances on the intermediate nucleon and
derive the corresponding coupling constants from the partial decay
widths. In section 4 we present numerical results for the contribution
to the binding energy of the trinucleons, which arises from these
interactions using oscillator and Malfliet-Tjon \cite{Malfliet}
wavefunctions as well as Paris and Bonn OBEPQ \cite{Stadler,Alfred}
wavefunctions.
An estimate that indicates that the $D$-state component
of the trinucleon bound state contributes only an insignificant
contribution to the net matrix element of the $\pi$-short range
three-nucleon interaction is presented in section 5. Finally section 6
contains a concluding discussion. The partial wave decomposition
of the TNI is described in the Appendix.\\

\newpage

\centerline{\bf 2. Pion-scalar and -vector meson exchange
three-nucleon interactions}
\vspace{0.5cm}

Realistic models for the nucleon-nucleon interaction as eg. the Bonn
\cite{Bonntab5}, Nijmegen \cite{Nijmegen}, and Paris \cite{Paris} models
are based on meson
exchange models, in which the Lorentz invariant on-shell nucleon-nucleon
scattering amplitude is constructed from phenomenological meson-nucleon
Lagrangians, and then used as an off-shell kernel in a wave-equation. In
practice a nonrelativistic ``adiabatic" approximation is also involved
\cite {Charap,Chemtob}. The important point is that a given model for
the on-shell amplitude that is extrapolated off shell in this way
automatically implies a model for the $NN\rightarrow NNN\bar N$ amplitude
as
well. This amplitude is the central component in the three-nucleon
interactions that arise from excitation of virtual nucleon-antinucleon
pairs on the intermediate nucleon (Fig. 1a). Thus any meson exchange model
for the nucleon-nucleon interaction will by construction imply the
presence of three-nucleon interactions of this type, which formally can
be derived directly from the $NN$ interaction model without any need for
further assumptions. \\

We shall here consider the $\pi$-scalar and $\pi$-vector meson
exchange TNI's of this form. An important part of the motivation for
this is the recent observation that the off shell $\pi+N\rightarrow
N+$scalar and $\pi +N\rightarrow N+$vector amplitudes are ``almost"
observable, in that they successfully describe most of the cross section
for the reaction $pp\rightarrow pp\pi^0$ near threshold \cite{pppi}.
Thus the
theoretical model for these amplitudes can be viewed as having a good
empirical foundation.
In the derivation of these three-nucleon amplitudes we
shall use the usual phenomenological scalar (``$\sigma$", ``$a_0$")
and vector meson (``$\omega$", ``$\rho$") Lagrangians, but in the end
will replace the pure boson exchange interactions by the corresponding
``effective" scalar and vector meson exchange interactions that form
the most important short range components of the nucleon-nucleon
interaction.\\

To construct the $\pi$-scalar (``$\sigma$")-meson three-nucleon
interaction that corresponds to the Feynman diagram in Fig. 1a we
employ the $\pi NN$ and $\sigma NN$ couplings

$${\cal L}_{\pi NN}
=i{f_{\pi NN}\over m_\pi}\bar \psi \gamma_5 \gamma_\mu
\partial_\mu \vec \phi_\pi \cdot \vec \tau \psi,\eqno(2.1a)$$

$${\cal L}_{\sigma NN}=g_\sigma \bar \psi \phi_\sigma \psi.\eqno(2.1b)$$
Here $\vec \phi_\pi$ is the isovector pion and $\phi_\sigma$ the
isoscalar scalar meson field and $f_{\pi NN}$ and $g_\sigma$ the
corresponding coupling constants. \\

The three-nucleon interaction that corresponds to the Feynman diagram
in Fig. 1a is obtained by retaining only the negative energy part of
the fermion propagator for the intermediate nucleon and adding the
term with the pion and scalar meson couplings in reversed order.
In the actual construction of the operator we exploit the fact that
the
nucleons are nearly on shell, and use the Dirac equation to
simplify the algebra. In this way a contact term operator arises,
which has to be retained along with the pair term in the three-
nucleon interaction operator.
The
resulting three-nucleon interaction has the simple form

$$V_{\pi \sigma}={g_\sigma^2\over m_N}({f_{\pi NN}\over
m_\pi})^2\frac{\vec \sigma^1\cdot \vec k_\pi \,\, \vec \sigma^2\cdot
\vec k_\sigma}{(k_\pi^2+m_\pi^2)(k_\sigma^2+m_\sigma^2)}\vec \tau^1
\cdot \vec \tau^2+({\rm permutations}).\eqno(2.2)$$
Here the direction of the meson momenta are taken so that they point
away from the intermediate nucleon,
that is indicated by the superscript 2 on the spin
and isospin operators. The symbol ``permutations"
stands for first adding a term, in which the nucleon coordinates 1
and 2 are exchanged and then taking into account the additional (4)
terms in which the intermediate nucleon-nucleon pair is excited on
nucleon 1 and 3 in turn.\\

In order to make this expression for the pion-scalar TNI consistent
with a realistic model for the nucleon-nucleon interaction it is
natural to replace the term $-g_\sigma^2/(k_\sigma^2+m_\sigma^2)$ in
the expression (2.2) by the corresponding general ``effective" isospin
independent scalar component $v_S^+(\vec k_\sigma)$ of the
nucleon-nucleon interaction, which may be constructed using the method
of ref. \cite{Blunden}. In a similar way it is natural to replace the term
$(f_{\pi NN}/m_\pi)^2/(k_\pi^2+m_\pi^2)$ by the effective isospin
dependent pseudoscalar exchange potential
$v_P^-(\vec k_\pi)/4m_N^2$ of the
nucleon-nucleon interaction. In this way the short range modifications
- i.e. form factors - of the simple meson exchange interactions are
determined by the corresponding components of the nucleon-nucleon
interaction model.  \\

Specifically, the method of ref. \cite{Blunden} rewrites a nonrelativistic
nucleon-nucleon potential model on the energy shell in terms of five
nonrelativistic spin amplitudes which can be viewed as nonrelativistic
limits of five relativistic Fermi invariants.  For the construction of
the TNI's in this paper we need the isospin independent ($^+$) and isospin
dependent ($^-$) Fermi invariants: scalar ($S$), pseudoscalar ($P$), and
vector ($V$).  The Fermi invariant potential coefficients
$v^{\pm}_j(\vec{k})$,  $j=A,P,V$ are obtained as linear combinations of
the nonrelativistic components of a given potential. The Fermi invariant
potential coefficients $v^{\pm}_j$ are functions of $k^2$ only which
means that the underlying interactions have no energy dependence. The
procedure and results for carrying out this program for a potential
(such as Paris), which has a short range
behavior determined in coordinate space, are displayed in \cite{Blunden}.
 If the potential is already expressed in terms of relativistic invariants
corresponding to the exchange of scalar, vector, and pseudoscalar bosons
(such as the Bonn or Nijmegen potentials) the procedure amounts to the
replacement (for example)

$$\frac{(f_{\pi NN}/m_\pi)^2} {m_\pi^2 + k_\pi^2}\rightarrow
\frac{v_P^-(\vec k_\pi)}{4m_N^2}
\rightarrow {(f_{\pi NN}/m_\pi)^2\over
k_\pi^2+m_\pi^2}({\Lambda_\pi^2-m_\pi^2 \over \Lambda_\pi^2+k_\pi^2}
)^2, \eqno(2.2a)$$
 where the meson-nucleon-nucleon vertex
$(\Lambda_\pi^2-m_\pi^2)/(\Lambda_\pi^2+k_\pi^2)$, is of the form chosen
for the Bonn OBEPQ potential used in our numerical investigations.
In either case, the short range behavior of the
TNI so constructed is fully determined by the short range behavior of the
corresponding nucleon-nucleon interaction, which is the issue at
hand. \\

In the case of an isospin-1 scalar meson exchange ($a_0$-channel) the
TNI that corresponds to the expression (2.2) is

$$V_{\pi a}={g_a^2\over m_N}({f_{\pi NN}\over m_\pi})^2\frac{\vec
\sigma^1 \cdot \vec k_\pi}{(k_\pi^2+m_\pi^2)(k_a^2+m_a^2)}$$
$$\{\vec \sigma^2\cdot \vec k_a\vec \tau^1\cdot \vec \tau^3+2i\vec
\sigma^2\cdot \vec P_2\vec \tau^1\cdot \vec \tau^2 \times \vec
\tau^3\}.\eqno(2.3)$$
Here $g_a$ is the $a_0 NN$ coupling constant and $m_a$ and $\vec k_a$
the mass and momentum of the exchanged $a_0$ respectively. In view of
the smallness of the nucleon momenta in the bound states we shall not
consider the
non-local
term in this TNI, which contains the momentum $\vec
P_2=(\vec p_2+\vec p\,'_2)$ of the intermediate nucleon. In order to
make the $\pi a_0$ TNI (2.3) consistent with the models for the
nucleon-nucleon interaction we shall replace the simple $a_0$ exchange
interaction in (2.3) by the corresponding isospin dependent scalar
exchange component of the nucleon-nucleon interaction: $g_a^2/
(m_a^2+k_a^2)\rightarrow -v_S^-(\vec k_a)$.\\

To construct the $\pi \omega$ TNI we employ the $\omega NN$ Lagrangian

$${\cal L}=ig_\omega \bar \psi \gamma_\mu \omega_\mu \psi,\eqno(2.4)$$
where $g_\omega$ is the $\omega NN$ coupling constant in addition to
(2.1a). The resulting TNI potential is

$$V_{\pi\omega}=-{g_\omega^2\over m_N}({f_{\pi NN}\over
m_\pi})^2\frac{\vec \sigma^1 \cdot \vec k_\pi \,\, \vec \sigma^2 \cdot
\vec k_\omega}{(k_\pi^2+m_\pi^2)(k_\omega^2+m_\omega^2)}\vec
\tau^1\cdot \vec \tau^2 + ({\rm permutations}).\eqno(2.5)$$
This interaction, which arises from the charge component of the
$\omega$-field is similar in form to the $\pi \sigma$ TNI (2.2), but
has the opposite sign. As in the case of the two-nucleon interaction
there will therefore be a strong partial cancellation between the
$\pi\sigma$ and $\pi\omega$ three-nucleon interactions, such that the
$\pi\sigma$ typically is the stronger interaction, because of the
somewhat longer (intermediate) range of the effective scalar meson
exchange interaction.\\

The $\pi \rho$ exchange TNI has one component that arises from the
charge component and one that arises from the spatial component. The
first one of these has the form

$$V_{\pi\rho}^C=-{g_\rho^2\over m_N}({f_{\pi NN}\over
m_\pi})^2\frac{\vec \sigma^1\cdot \vec k_\pi \,\, \vec \sigma^2\cdot
\vec k_\rho}{(k_\pi^2+m_\pi^2)(k_\rho^2+m_\rho^2)}\vec \tau^1\cdot
\vec \tau^3,\eqno(2.6)$$
and the expression for the latter is

$$V_{\pi\rho}^S=-i{g_\rho^2\over m_N}({f_{\pi NN}\over
m_\pi})^2\frac{\vec \sigma^1\cdot \vec k_\pi}{(k_\pi^2+m_\pi^2)
(k_\rho^2+m_\rho^2)}$$
$$\sigma^2\cdot [2 \vec P_3+i\vec \sigma^3\times \vec k_\rho]\vec
\tau^1 \cdot \vec \tau^2 \times \vec \tau^3.\eqno(2.7)$$
The local part of the $\pi\rho$ exchange three nucleon interaction
$V_{\pi\rho}^S$ is referred to as the ``$\pi\rho$ Kroll-Rudermann
interaction" \cite{TM2} or the ``seagull" \cite{Robilotta} in the
literature, and has been considered in the trinucleon before
\cite{Stadler,Sasrho,Pena}.  In the treatment of the Tucson-Melbourne
TNI (an expansion of the $\rho N \rightarrow \pi N$ amplitude)
$V_{\pi\rho}^C$ is small because of a near cancellation between the
intermediate nucleon-antinucleon part and the $\rho$ analog
Fubini-Furlan-Rossetti contribution to pion photoproduction.  The
isoscalar $V_{\pi\rho}^C$ corresponds to the remaining lead term in
eqs. 2.13b and 2.14a of [4a]. It was estimated in nuclear matter in [4a],
derived and then neglected altogether in the Brazil TNI
\cite{Robilotta}, and is evaluated in the trinucleon for the first time
in the present study. To make the present expressions for the
$\pi\omega$ and $\pi\rho$ exchange three-nucleon interactions consistent
with the nucleon-nucleon interaction model we shall replace the bare
vector meson interactions $g_\omega^2/(k_\omega^2+m_\omega^2)$ and
$g_\rho^2/(k_\rho^2+m_\rho^2)$ with the corresponding isospin
independent $(v_V^+(\vec k_\omega))$ and isospin dependent $v_V^-(\vec
k_\rho))$ vector exchange components of the nucleon interaction as
suggested in ref. \cite{Blunden}.

\vspace{1cm}

\centerline{\bf 3. Intermediate ${\bf N(1440)}$ resonances}
\vspace{0.5cm}

Excitation of virtual $N(1440)$ (Roper) resonances on the intermediate
nucleon also contributes a weak but still significant $\pi$-scalar and
$\pi$-vector meson exchange three nucleon interaction (Fig. 1b).
Naturally a contribution to the $\pi\pi$ three-nucleon interaction
also arises from intermediate $N(1440)$ excitation, but this is
effectively included in the Tucson-Melbourne TNI, which is based on an
off shell extrapolation of the complete $\pi N$ scattering
amplitude.\\

To construct the $\pi\sigma$ and $\pi\omega$ three nucleon
interactions that are associated with intermediate $N(1440)$ resonance
excitation we employ the effective Lagrangians

$${\cal L}_{\pi NN^*}=i{f^*_\pi\over m_\pi}\bar \psi_* \gamma_5 \gamma_\mu
\partial_\mu \vec\phi \cdot \vec\tau \phi+{\rm h.c.},\eqno(3.1a)$$

$${\cal L}_{\sigma NN^*}=g_\sigma^* \bar \psi_* \phi_\sigma \psi
+{\rm h.c.},\eqno(3.1b)$$

$${\cal L}_{\omega NN^*}
=ig_\omega^*\bar \psi_* \gamma_\mu \omega_\mu \psi
+{\rm h.c.}.\eqno(3.1c)$$
Here $\psi_*$ denotes the Roper resonance spinor field and
$f^*_\pi,\,\,g_\sigma^*$ and $g_\omega^*$ are the $\pi NN^*$, $\sigma
NN^*$ and $\omega NN^*$ coupling strengths, $N^*$ being an
abbreviation for the $N(1440)$. The expressions for the $\pi\sigma$
and $\pi \omega$ TNI potentials that are associated with intermediate
$N(1440)$ excitation can then be derived in a straightforward way, the
results being

$$V_{\pi\sigma}^*=-{2g_\sigma g_\sigma^*\over m^*-m_N}{f_{\pi
NN}f^*_\pi\over m_\pi^2}\frac{\vec \sigma^1\cdot \vec k_\pi \,\, \vec
\sigma^2\cdot \vec
k_\pi}{(k_\pi^2+m_\pi^2)(k_\sigma^2+m_\sigma^2)}\vec \tau^1\cdot
\vec \tau^2 +({\rm permutations}),\eqno(3.2a)$$

$$V_{\pi\omega}^*={2g_\omega g_\omega^* \over m^*-m_N}{f_{\pi
NN}f^*_\pi\over m_\pi^2}\frac{\vec \sigma^1\cdot \vec k_\pi \,\, \vec
\sigma^2\cdot \vec
k_\pi}{(k_\pi^2+m_\pi^2)(k_\omega^2+m_\omega^2)}(\vec \tau^1\cdot
\vec \tau^2).\eqno(3.2b)$$
Here $m^*$ is the mass of the $N(1440)$ resonance.
It is worth noting that the terms that depend on the pion
momentum $\vec k_\pi$ in these three-nucleon
interactions have the same form as the one-pion exchange
interaction between two nucleons.\\

The $\pi NN^*$ coupling constant $f^*_\pi$ may be calculated from the
$N^*\rightarrow N\pi$ partial decay width as

$${f^{*2}_\pi\over 4\pi}=\frac{m^*
m_\pi}{(m^*+m_N)p(E_N-m_N)}\Gamma(N^*\rightarrow N\pi).\eqno(3.3)$$
Here $p$ is the nucleon momentum and $E_N$ the nucleon energy in the
rest frame of the decaying $N^*$. With the $N\pi$ branching ratio of
the total decay width 350 MeV being 60\% we obtain $f^*_\pi/4\pi \simeq
0.031$, which is somewhat less than one half of the corresponding
value $f_{\pi NN}^2/4\pi\simeq 0.08$ for the $\pi NN$ coupling
strength.\\

The determination of the $\sigma NN^*$ and $\omega NN^*$ coupling
strength is associated with considerably larger uncertainties. To
obtain an estimate for the $\sigma NN^*$ coupling constant we
calculate it from the decay width for $N^* \rightarrow
N(\pi\pi)_{S-{\rm wave}}^{I=0}$ as

$${g_\sigma^{*2}\over 4\pi}={m^*\over p(E_N+m_N)}\Gamma(N^*\rightarrow
N(\pi\pi)_{S-{\rm wave}}^{I=0}).\eqno(3.4)$$
Here we assume that all of the $I=0$ $S$-wave part of the $\pi\pi$
continuum can be interpreted as a broad effective $\sigma$-meson. The
branching ratio for this decay channel is 5-15\%. Assuming $m_\sigma$
= 410 MeV at the midpoint between the $\pi\pi$ threshold and
kinematical phase space cutoff we obtain

$${g_\sigma^{*2}\over 4\pi}\simeq 0.1 \eqno(3.5)$$
for $\Gamma(N^*\rightarrow N(\pi\pi)_{S-{\rm wave}}^{I=0}$) = 35 MeV.
This value for $g_\sigma^{*2}/4\pi$ is expected to have an uncertainty of
about a factor 2.\\

As the $N^*$ cannot decay into a $N\omega$ state, the coupling constant
$g_\omega$ cannot be determined directly from empirical data. We shall
here assume that $g^*_\omega/g_\omega=g_\sigma^*/g_\sigma$ as suggested
by the constituent quark model. In the Bonn boson exchange model OBEPQ for
the nucleon-nucleon interaction this ratio is 1.55 \cite{Bonntab5}. This would
then suggest that

$${g_\omega^{*2} \over 4\pi}=0.24,\eqno(3.6)$$
a value with which a substantial uncertainty margin also has to be
associated.\\

The bare meson exchange potentials in the TNI's (3.2a) and (3.2b) will
be modified at high values of momentum transfer by shorter range dynamics
in the same way as the $NN$ interaction. To describe this short range
modification in a way that is consistent with that of the $NN$
interaction we shall introduce the same vertex factors as in the Bonn
boson exchange model OBEPQ for the NN interaction \cite{Bonntab5} by
means of the substitutions

$${1\over m_\pi^2 + k_\pi^2}\rightarrow {1\over
k_\pi^2+m_\pi^2}({\Lambda_\pi^2-m_\pi^2 \over \Lambda_\pi^2+k_\pi^2}
)^2,\eqno(3.7a)$$

$${1\over m_\sigma^2+k_\sigma^2}\rightarrow {1\over
k_\sigma^2+m_\sigma^2}({\Lambda_\sigma^2-m_\sigma^2\over
\Lambda_\sigma^2+k_\sigma^2})^2,\eqno(3.7b)$$

$${1\over m_\omega^2+k_\omega^2}\rightarrow {1\over
k_\omega^2+m_\omega^2}({\Lambda_\omega^2-m_\omega^2\over
\Lambda_\omega^2+k_\omega^2})^2.\eqno(3.8b)$$
For the form factor mass scale parameters we use the values
$\Lambda_\pi$ = 1.3 GeV/$c^2$, $\Lambda_\sigma = \Lambda_\omega$ = 2.0
GeV/$c^2$.

\vspace{1cm}
\newpage

\centerline{\bf 4. Numerical estimates}
\vspace{0.5cm}

We shall here estimate the contribution of the $\pi$-scalar and
$\pi$-vector exchange TNI to the binding energy of the trinucleons
with a pure $S$-state wavefunction model. In the first estimates
presented, the orbital part of the
wavefunction is described by the harmonic oscillator and three-channel
Malfliet-Tjon I-III model wavefunction \cite{Malfliet}. The use of the harmonic
oscillator model is motivated by the fact that the resulting matrix
elements can be reduced to quadrature of very simple expressions,
which allows the qualitative features to be illuminated. The numerical
values for the matrix elements
of the three-nucleon interactions considered here that are
obtained with the harmonic oscillator
and Malfliet-Tjon wave function models turn out to be very similar.\\

In the case of a wavefunction with only a completely symmetric
$S$-state component the radial matrix element may be
expressed as

$$<V_3>=\int {d^3 \tau\over (2\pi)^3}\int {d^3 v\over (2\pi)^3}g(\vec
\tau,\vec v)\phi_0^\dagger V_3(\vec \tau,\vec v)\phi_0.\eqno(4.1)$$
Here $\phi_0$ is the totally antisymmetric spin-isospin vector and
$\vec \tau$ and $\vec v$ are differences of nucleon Jacobi coordinates,
which are
related to the meson momenta in the three-nucleon interactions as

$$\vec k_\pi=(\vec \tau +{\vec v\over 2}),\;\;\;\vec k_{\sigma,\omega}=\vec
v.\eqno(4.2)$$
In (4.1) the function
$g(\vec \tau,\vec v)$ is the Fourier transform of the
nucleon density function in coordinate space, which in the
case of the harmonic oscillator model takes the form

$$g(\vec \tau,\vec
v)=e^{-\tau^2/2\alpha^2}e^{-3v^2/8\alpha^2},\eqno(4.3)$$
$\alpha$ being the oscillator parameter $\sqrt{m\omega_0}$ for which
we use the value 0.60 $\rm fm ^{-1}$.\\

The oscillator model leads to the following expression for the matrix
element of the $\pi\sigma$ three-nucleon interaction (2.2) in $^3H$
and $^3He$:

$$<V_{\pi\sigma}>={3\over 2\pi^4}{g_\sigma^2\over m_N}({f_{\pi
NN}\over m_\pi})^2\int_{0}^{\infty}dv  e^{-v^2/2\alpha^2}{v^3\over
v^2+m_\sigma^2}$$
$$\int_{0}^{\infty} dk e^{-k^2/2\alpha^2}{k^3\over
k^2+m_\pi^2}\sqrt{{\pi \alpha^2 \over kv}} I_{3/2}({kv\over
2\alpha^2}),\eqno(4.4)$$
where $I_{3/2}$ is a modified Bessel function. The corresponding
expression for the matrix element of the $\pi a_0$ three-nucleon
interaction (2.3) is obtained by the substitutions
$g_\sigma\rightarrow g_{a}$ and $m_\sigma\rightarrow m_a$ and an
overall change of sign in the expression (4.4), if the terms
proportional to the total momentum of the intermediate nucleon in
(2.3) are neglected. The sign change arises from different isospin
dependence in (2.2) and (2.3).\\

The similarity in form between the $\pi\sigma$ (2.2) and $\pi \omega$ (2.5)
three-nucleon interactions makes it obvious that the expression for
the matrix element of $V_{\pi \omega}$ is given by an expression of
the form (4.4) with $g_\sigma$, $m_\sigma$ replaced by $g_\omega$ and
$m_\omega$, and with an overall minus sign. The expression for the
matrix element of the $\pi\rho$ three-nucleon interaction $V_{\pi
\rho}^C$ (2.6) is obtained by the corresponding substitution in (4.4)
of $g_\sigma \rightarrow g_\omega$, $m_\sigma \rightarrow m_\rho$, but
without any change of the overall sign.\\

With the oscillator model density function (4.3) the matrix element of
the $N(1440)$ intermediate state $\pi\sigma$ three-nucleon interaction
(3.2a) takes the form

$$<V_{\pi\sigma}^*>={3\over \pi^4}{g_\sigma g_\sigma^*\over m^*
-m_N}{f_{\pi NN}f_\pi^*\over
m_\pi^2}\int_{0}^{\infty}dve^{-v^2/2\alpha^2}{v^2\over
v^2+m_\sigma^2}$$
$$\int_{0}^{\infty}dke^{-k^2/2\alpha^2}{k^4\over
k^2+m_\pi^2}\sqrt{{\pi \alpha \over kv}}I_{1/2}({kv\over
2\alpha^2}).\eqno(4.5)$$
The corresponding expression for the matrix element of the $\pi\omega$
exchange three-nucleon interaction $V_{\pi\omega}^{*}$ (3.2b) can
be obtained simply from (4.5) by means of the substitutions
$g_\sigma^*\rightarrow g_a^*$ and $m_\sigma\rightarrow m_\omega$ and
an accompanying overall sign change. \\

The harmonic oscillator model of a three-body bound state has special
properties which allow a practical implementation of eq. (4.1) as shown
above.  Other techniques must be used for more realistic wave functions
because the triton wavefunction depends on the initial and final Jacobi
variables of the three nucleons.  In general, the momentum transfer
differences of eq. (4.2) must be supplemented by sums of Jacobi variables
which, for the harmonic oscillator, cancel the norm integral and
therefore need not be considered in eq. (4.1).
Further numerical calculations were made according to the methods
described in \cite{Pena} and [3b].  The  $S$-state wavefunction that
corresponds to the Malfliet-Tjon I-III interaction is given on a mesh
over the coordinate space nucleon Jacobi variables described in
\cite{Malfliet}.  After a Fourier transformation of the potentials of
sections 2 and 3, the three dimensional integrals over two vector
variables are straightforward to carry out numerically.  On the other
hand, in order to
employ contemporary three-body wavefunctions from momentum space Faddeev
calculations it is necessary to make a partial wave decomposition of
$V_3$ as was first done in [3b].  The final result of the partial wave
decomposition of the potentials developed here is presented in Appendix
A. \\

In Table I we give the numerical values for the matrix elements of the
three-nucleon $\pi$-short range three-nucleon interactions that are
associated with excitation of intermediate nucleon-antinucleon pairs.
The
numerical values have been given both for the case of the
oscillator
model wavefunction and the $S$-state wavefunction that
corresponds to the Malfliet-Tjon I-III interaction \cite{Malfliet}.
The numbers in the
table correspond to the case when the Bonn boson exchange potential
model OBEPQ (Table V, \cite{Bonntab5}) parameterization has been used to
construct the ``effective" pion (pseudoscalar) and scalar and vector
exchange potentials as described in section 2 above.\\

In Table II we give also for this parameterization
of the TNI the numerical values for the Bonn OBEPQ and Paris
wavefunction models --- see Appendix for the technicalities involved
--- corresponding to the
the $\pi$-$\omega$ and $\pi$-$\sigma$ exchange
force only (according to Table I they dominate the overall $\pi$-scalar
and $\pi$-vector exchange TNI).
Table II confirms the trend of repulsion obtained with the more
schematic wavefunction models used in the calculations of Table I.
It also stresses once more the traditional extreme results of the
OBEPQ potential for the triton binding energy.\\

All wavefunction models indicate that the $S$-state matrix elements of
the $\pi$-short range three-nucleon interactions that are associated
with intermediate $N\bar N$ pair excitation are about 100 keV-200 keV
(99 keV, 126 keV, 134 keV, 194 keV respectively for the
H.O., MT, Paris and Bonn model wavefunctions), when these three nucleon
interactions are
constructed so as to be consistent with the Bonn OBEPQ model for the $NN$
interaction. This 100 keV has a substantial theoretical uncertainty
margin that is due to the remaining uncertainty in the short range
behaviour of the nucleon-nucleon interaction. To illustrate this we
have also constructed these $\pi$-scalar and $\pi$-vector exchange
three-nucleon interactions from the Paris model for the
nucleon-nucleon interaction \cite{Paris} . In that potential model the isospin
independent scalar exchange component is much weaker than in other
realistic phenomenological potential models, and is in fact repulsive
at short range \cite{Blunden}. As a consequence that matrix element
$<V_{\pi\sigma}>$ calculated with the oscillator wave function model
is only 143 keV (or 113 keV for a even more consistent calculation with
the Paris wavefunction) as compared to the corresponding value 392 keV
obtained with the Bonn OBEPQ potential. This reduction of $<V_{\pi\sigma}>$
makes the net matrix element of the $\pi$-short range three nucleon
attractive, when constructed from the parametrized
Paris potential ($\simeq$ --200
keV for the oscillator wavefunction, or $\simeq$ --101 keV for the Paris
wavefunction).\\

The difference of almost 300 keV (and a change of sign)
between these two most consistent calculations presented here (a Bonn
wavefunction and Bonn TNI compared to the Paris wavefunction and Paris
TNI) may also in part be due to an inconsistency in the $\mu$ parameter
of the latter combination.  The value of $\mu$ labels an ambiguity in
the relativistic corrections ( ie from $v/c$-expansion methods) to
operators involving pion exchange \cite{Friar}.  The same continuous
parameter also acts as a chiral rotation to determine the pseudoscalar
($\mu$ = 0) or pseudovector ($\mu$ = 1) content of the pion-nucleon
coupling in models of the nucleon-nucleon interaction or TNI's which
obey approximate chiral symmetry \cite{CoonFriar}.  Fixing arbitrarily
this parameter (or the more general parameter $\tilde {\mu}$ which
includes a specification of energy transfer in the non-relativistic $\pi
NN$ vertex \cite{Adam}) means choosing a special representation of a
unitarily equivalent class of operators. Because of this unitary
equivalence, the observables should not depend on the value of $\tilde
{\mu}$, i.e. on the choice of representation, provided that all
operators and wave functions are chosen consistently. The derivation of
$V_{\pi\sigma}$ of equation (2.2) and of $V_{\pi\omega}$ of equation
(2.5) corresponds to $\tilde{\mu}$ = -1, as does the one-pion-exchange
part of the Bonn OBEPQ. Thus it is completely consistent to determine
the short range modifications of the simple meson exchange interactions
of eqs. (2.2) and (2.5) by the corresponding components of the Bonn
OBEPQ nucleon-nucleon interaction model.  The same replacement procedure
with the Paris potential is, however, more problematic, as the
parameter of that potential has been identified to be $\tilde{\mu}$ = 0
\cite{Adam}.  This inconsistency between the derivation of the TNI and
the choice of short range modification is a possible cause of the
discrepancy between our ``consistent" calculations.  Note that the
harmonic oscillator NN model and the Malfliet-Tjon NN model do not have
any pion exchange component so this particular consistency question does
not arise. Also, the other parameter $\nu$ of the unitarily linked
operators of this paper is always $\nu= 1/2$, corresponding to no meson
retardation, so no inconsistency can be attributed to $\nu$. \\

The TNI's due to an $N(1440)$ intermediate state have much smaller
binding energy effects than those just discussed, a fortunate result for
our goal of building TNI's consistent with a given nucleon-nucleon
interaction. The matrix element of the $N(1440)$ intermediate state
$\pi\sigma$ three-nucleon interaction $V_{\pi\sigma}^*$ (4.5) for the
$S$-state oscillator model for the bound trinucleon states is 200 keV.
Here we have used the parameters $m_\sigma$ = 530 MeV and
$g_\sigma^2/4\pi$ = 8.2797 for the mass and (nucleon) coupling constants
for the exchanged $\sigma$ meson as suggested by the Bonn OBEPQ
potential \cite{Bonntab5}, and the short range form factors (3.7). The
values of the meson$-NN^*$ coupling constants are those derived in
section 3. The corresponding matrix element of the $\pi\omega$ TNI
$V_{\pi\omega}^*$ (3.2b) is --150 keV so that the net contribution of the
TNI associated with the $N(1440)$ is repulsive and $\simeq$ 50 keV. This
is about one half as large as the repulsive contribution that is due to
the TNI associated with intermediate $N\bar N$ pair excitation.  When
the same calculations are made with the substitution of the
Malfliet-Tjon wave function for the oscillator wavefunction, the
individual contributions are  $\simeq$ 65 keV from $V_{\pi\sigma}^*$
and   $\simeq$ --45 keV from $V_{\pi\omega}^*$, for a total repulsion of
$\simeq$ 20  keV.  This total is again small compared to the repulsive
contribution of $\simeq$ 100  keV that is due to the TNI associated with
intermediate $N\bar N$ pair excitation.  Finally, in our most consistent
calculation (OBEQ wavefunction and the Bonn boson exchange
parametrization of the TNI),
the total repulsion from the TNI associated with the
$N(1440)$ is about one third (+0.068:+0.194 keV) of that of the
intermediate $N\bar N$ pair excitation with the Bonn OBEPQ triton
wavefunction. \\

In  the $S$-state oscillator trinucleon model and the semirealistic,
Malfliet-Tjon NN potential wave function and the more realistic Paris
potential,  the combined contribution of
all the $\pi$-short range TNI's considered here is thus about 150 keV,
which corresponds to a significant fraction of the repulsive TNI
contribution needed to explain the binding energies of the three-nucleon
bound states. For the Bonn OBEPQ triton wavefunction this estimate is
enhanced, as usual, to about 250 keV repulsion.  To put these results in
perspective we remind the reader that a similar evaluation of the
$\pi-\rho$ "Kroll-Ruderman interaction"
of Ref. [4b] with the Malfliet-Tjon
NN potential wave function yields about 340 keV repulsion when the
formfactors are chosen similar to those of this paper.\\

This estimate of the repulsive contribution due to the
$\pi$-short range three-nucleon interaction would a priori be expected
to be altered by the $D$-state terms (or tensor correlations)
in the trinucleon bound states. A suggestive estimate of these
alterations is provided by the comparison of the "Kroll-Ruderman"
interaction" used perturbatively with the simple Malfliet-Tjon NN
potential wave function in Ref. [4b] and the expectation values of the
same interaction with triton wave functions fully correlated by two- and
three-nucleon interactions from Table IV of Ref. [8c].  One finds a
reduction in the contribution of this nucleon-antinucleon pair term by
20\% to 40\% from the Malfliet-Tjon result when a careful calculation
with the Nijmegen or Paris triton wavefunction is made.  In the
following section we attempt to understand these comparisons by showing
(again with the aid of the harmonic oscillator trinucleon model) that
the effect of the $D$-state admixture  is a reduction of the pair term
and enhancement of the isobar term so that effect on the net repulsive
$\pi$-short range TNI is very small.

\vspace{1cm}
\newpage
\centerline{\bf 5. Estimate of the D-state contribution}
\vspace{0.5cm}

The wavefunctions of the bound trinucleons that are obtained with
realistic nucleon-nucleon interaction models contain $D$-state
admixtures of the order 10\%. As the $\pi$-short range three-nucleon
interactions derived in sections 2 and 3 above have a rank-2 spatial
tensor component they should be expected to have large $SD$-cross
term matrix elements. In order to obtain a numerical estimate of the
relative magnitude of this $SD$-state contribution we construct a
model $D$-state wavefunction as \cite{Gerjuoy}

$$\varphi_D=N_D\sum_{i<j}S_{ij}r_{ij}^2 \vec \tau_i \cdot \vec \tau_j
\varphi_0 \phi_0,\eqno(5.1)$$
where $\varphi_0$ is the oscillator model trinucleon orbital
wavefunction

$$\varphi_0=({\alpha^2\over \pi \sqrt{3}})^3
e^{-\alpha^2r^2/4}e^{-\alpha^2\rho^2/3}.\eqno(5.2)$$
Here $\vec r$ and $\vec \rho$ are differences of the initial
and final state nucleon Jacobi
coordinates. The corresponding $S$-state wavefunction is simply
$\varphi_S=\sqrt{P_S}\varphi_0$, where $P_S$ is the $S$-state
probability. If the three-body tensor terms as $S_{12} S_{23}$ are
dropped in the calculation of the $D$-state normalization factor $N_D$,
it has the value

$$N_D={\alpha^2\over 18 \sqrt{5}}\sqrt{P_D},\eqno(5.3)$$
where $P_D$ is the $D$-state probability.\\

As the main source of
the $D$-state admixture is the pion exchange
tensor interaction it is a natural approximation to evaluate the $SD$
cross term matrix elements of the three-nucleon interactions with only
the nucleon pair that involves a pion exchange having a $D$-state
component. The relevant $SD$-cross term density function is then

$$g_D^{12}(\vec \tau, \vec v)=-{\sqrt{P_D}\over 18\sqrt{5}}S_{12}(\vec
\tau) \vec \tau^1\cdot \vec \tau^2
e^{-\tau^2/2\alpha^2}e^{-3v^2/8\alpha^2},\eqno(5.4)$$
where the tensor operator $S_{12}(\vec\tau)$ is defined as

$$S_{12}(\vec \tau)=3(\vec \sigma^1 \cdot \hat\tau)(\vec \sigma^2\cdot
\hat \tau)-\vec \sigma^1\cdot \vec \sigma^2.\eqno(5.5)$$
With this model density function the $SD$-state matrix element of the
$\pi\sigma$ TNI (2.2) can be expressed in the form

$$<V_{\pi\sigma}>_{SD+DS}=-{\sqrt{P_DP_S}\over
3\pi^4\sqrt{5}}{g_\sigma^2\over m_N}({f_{\pi NN}\over m_\pi})^2$$
$$\int_0^\infty e^{-v^2/2\alpha^2}\frac{v^3}{v^2+m_\sigma^2}
\int_0^\infty dke^{-k^2/2\alpha^2}{k^3 \over
k^2+m_\pi^2}\sqrt{{\pi\alpha^2\over kv}}$$
$$\{5vk I_{1/2}({kv\over 2\alpha^2})-6[k^2+{v^2\over
4}]I_{3/2}({kv\over 2\alpha^2})
+vk I_{5/2}({kv\over 2\alpha^2})\}.\eqno(5.6)$$
The expressions for the $SD$ matrix element of the $\pi a_0$,
$\pi\omega$ and $\pi\rho$ exchange three nucleon interactions that are
associated with intermediate $N\bar N$ pair excitation can be
obtained by making the same substitutions in eq. (5.5) as described
after eq. (4.5).\\

Assuming a 10\% $D$-state admixture we find the numerical values for
the $SD$-state matrix elements calculated using the expression (5.5)
to be $-80$ keV in the case of the $\pi\sigma$ TNI and $-20$ keV when
the contributions of the $\pi\sigma$, $\pi a_0$, $\pi\rho$ and
$\pi\omega$ three nucleon interactions are combined and constructed
from the Bonn OBEPQ model for the nucleon-nucleon interaction \cite{Bonntab5}.
The
effect of the $SD$-state cross term matrix element is thus to reduce
the repulsive $\pi$-short range contribution to the TNI that is
associated with $N\bar N$ pair excitation by about 20\%.\\

The $SD$ state matrix element of the $\pi\sigma$ exchange interaction
(3.2a) that is due to the excitation of intermediate $N(1440)$
resonances can be derived by the same method using the $SD$-state
density function (5.3). The resulting expression is

$$<V_{\pi\sigma}^*>_{SD+DS}={\sqrt{P_DP_S}\over
6\pi^4\sqrt{5}}{g_\sigma g_\sigma^*\over m^*-m_N}{f_{\pi
NN}f_\pi^*\over m_\pi^2}$$
$$\int dve^{-v^2/2\alpha^2}{v^2\over v^2+m_\sigma^2}\int
dke^{-k^2/2\alpha^2}{k^4\over k^2+m_\pi^2}$$
$$\sqrt{{\pi \alpha^2\over kv}}\{[24k^2-{v^2\over 2}]I_{1/2}({kv\over
2\alpha^2})$$
$$-24kv I_{3/2}({kv\over 2\alpha^2})+6v^2I_{5/2}({kv\over
2\alpha^2})\}.\eqno(5.7)$$
The expression for the matrix element of the corresponding $\pi\omega$
three nucleon interaction (3.2b) is obtained from (5.6) by the
substitutions $g_\sigma,\,\, g_\sigma^*\rightarrow g_\omega,\,\,
g_\omega^*$ and $m_\sigma \rightarrow m_\omega$ and a change of the
overall sign. The numerical value for $<V_{\pi\sigma}^*>_{SD+DS}$ we
find to be 94 keV, which is about one half of the corresponding $SS$
matrix element. Addition of the matrix element of the $\pi\omega$
exchange interaction $<V_{\pi\omega}^*>$ reduces this value to 21 keV.
The net effect of the $D$-state is thus to increase the total matrix
element of the $\pi$-short range TNI that arises from excitation of
intermediate $N(1440)$ resonances by about 40\%.\\

As the numerical value for the $SD$ state matrix elements of the
$\pi$-short range three nucleon interactions that are due to
excitation of intermediate nucleon-antinucleon pairs and $N(1440)$
resonances are almost equal in magnitude and opposite in sign we
conclude that the importance of the $D$-state admixture in the bound
trinucleon states is very insignificant in this instance. Hence the
estimates in section 4 that were obtained with pure $S$-state
wavefunction models and which are very similar for the wavefunction
models considered should be viewed as fairly robust.

\vspace{1cm}

{\bf 6. Discussion}
\vspace{0.5cm}

The present results demonstrate that the pion-scalar and pion-vector
meson exchange three-nucleon interactions are important on the general
scale of three-nucleon interactions. At the level of precision attained
by calculations with
the present realistic semiphenomenological  nucleon-nucleon
interactions, which also contain three-nucleon interactions,
this TNI has to be included in the calculation of
nuclear binding energies. The repulsive contribution of this
$\pi$-short range exchange TNI appears able to explain most if not all
of the repulsion hitherto ascribed to the purely phenomenological
spin-independent TNI of short range, which was introduced to achieve
agreement with the empirical binding energies of the few-nucleon
systems [7].\\

The numerical values of the matrix elements of the $\pi$-short range
three-nucleon interactions presented here should be realistic in spite of
the employment of pure $S$-state models of the wavefunction of the
bound three-nucleon system. By considering a schematic model for the
$D$-state component it was shown that due to a number of cancellations
the net effect of the $D$-state component for these TNI's is small.\\

The most uncertain in magnitude of the three-nucleon interactions
considered here is that associated with the excitation of intermediate
$N(1440)$ resonances. The main uncertainty in this interaction is due
to the unknown $\omega NN^*(1440)$ coupling constant. Fortunately, this
interaction has smaller effects on the triton binding energy than the
interaction due to nucleon-antinucleon pair terms which can be directly
related to realistic NN interactions in the manner shown here.

\vspace{1cm}

\centerline{ \bf Acknowledgements}

The work of M.T.P. was supported in part by JNICT, under contract
No. PBIC/C/CEN/1094/92 and ``Contrato Plurianual'', that of S.A.C.
under NSF grant PHY-94081347 and that of D.O.R by Academy of
Finland grant 7635. D.O.R thanks the Institute for Nuclear
Theory, Seattle for hospitality at the time this work was
completed.
We thank Alfred
Stadler for providing the Bonn
OBEPQ and Paris three-body wavefunctions and the
Iowa-Los Alamos three-body group for the Malfliet-Tjon wavefunctions.

\newpage

\centerline{\bf Appendix --- Partial wave decomposition }
\vspace{0.5cm}

While the calculations that employ the
semi-realistic
Gaussian and Malfliet-Tjon wave functions provide
results that
are qualitatively indicative,
present
state-of-the-art
three-nucleon calculations have provided
several examples of delicate cancellations that
may be missed by schematic wave function models.
This motivated a calculation based on
more realistic wavefunctions, such as the ones obtained with the
Paris and Bonn OBEPQ potentials (in references \cite{Stadler,Alfred}). These
two potentials
lead to quite different answers for the binding energy
of the triton, and
different probabilities for the S and D state components of
the wavefunction.\\

The expectation value of the TNI can be calculated by means of
equation (4.1). Although this equation is general,
it requires the knowledge of $g(\vec {\tau},\vec v)$.
Since this last function is not directly supplied by any standard
Faddeev code, it
demands extra computational effort.
The algorithm that is usually employed with realistic interaction
models is to evaluate
the matrix element
of the TNI in the partial wave decomposed basis used
in the wave function calculation. The partial
wave decomposition has been
presented in a fairly general way in ref. [3b].
We here review its main steps, keeping the notation close to
the one introduced in that work.\\

In the $LS$ coupling scheme the notation for the partial
wave decomposed wavefunction is
$$|pq \alpha \rangle_2=
|pq, [(l \lambda)L (s 1/2)S]JJ_z, (t 1/2)T T_z \rangle_2$$
where the index $2$ specifies the particle that is taken to be
the spectator in the definition of ($\vec p$, $\vec q$).
To calculate
the TNI matrix element in this basis,
we start by  separating
explicitly the spin
dependence of the TNI from the orbital one.
For this purpose
the TNI is decomposed in
spherical components and consequently in
spherical harmonics.
Subsequently closed form quadrature
over the angular arguments of these spherical
harmonics is performed. This requires expansion in
Legendre polynomials of the angle dependence of
the form factors and propagators. These expansions (in three angles)
are done numerically, through a Gaussian mesh of at least 20 points
for each angle.\\

To illustrate the technique we consider
the intermediate
$N\bar N$ $\pi-\sigma$ TNI.
For this particular case we have
(leaving out the trivial isospin
dependence),

$$ ({\vec {\sigma}}^2 \cdot {\vec {k_{\sigma}}})
   ({\vec {\sigma}}^1 \cdot {\vec {k_{\pi}}})=
\frac{4\pi}{3} k_{\sigma} k_{\pi} \sum_k\sqrt{2k+1}
\left[\sigma(1) \times \sigma(2)
\right]^k Y_{1 1}^{k} (\hat k_{\pi},\hat k_{\sigma}).
\eqno{(A.1)}$$
Here the
$Y_{1 1}^{k} (\hat k_{\pi}, \hat k_{\sigma})$ function stands
for the two (coupled) spherical
harmonics, which depend separately
on the momenta of each meson, i.e.,

$$\left[Y^{1} \times Y^{1} \right]^k= \sum_{m_1,m_2}
C^{1 1 k}_{m_1 m_2 m}
Y_{m_1}^{1} (\hat k_{\pi}) Y_{m_2}^{1} (\hat k_{\sigma}). \eqno{(A.2)}$$
After expressing  the exchanged meson momenta in terms of the Jacobi
coordinates $\vec p, \vec q$ (for the initial three-nucleon state) and
$\vec p\,', \vec q\,'$ (for the final three-nucleon state),
$$\vec {k}_{\pi} =(\vec p - \vec p\,') -\frac{1}{2} (\vec q - \vec q\,'),$$
$$\vec {k}_{\sigma} =(\vec p - \vec p\,') +\frac{1}{2} (\vec q - \vec q\,').$$
the
function $Y_{1 1}^k (\hat k_{\pi}, \hat k_{\sigma}) $ can be decomposed in
coupled spherical harmonics
of simpler arguments:
\begin{eqnarray*}
\lefteqn{
Y_{1 1}^k (k_{\pi},k_{\sigma})=
3 \sum_{r_1+r_2=1} \sum_{s_1+s_2=1}
 (-1)^{r_2} F(1,r_1,r_2) F(1,s_1,s_2) (\frac{1}{2}     )^{r_2+s_2}
}
\nonumber\\
& & \times
\frac{|\vec p -\vec p\,'|^{r_1+s_1}  |\vec q -\vec q\,'|^{r_2+s_2}}
{k_{\pi}k_{\sigma}}
\sum_{t_1 t_2}
\left\{
\begin{array}{ccc}
r_1 & r_2 & 1 \\
s_1 & s_2 & 1 \\
t_1 & t_2 & k_1
\end{array}
\right\}
C^{r_1 s_1 t_1}_{0 0 0}
C^{r_2 s_2 t_2}_{0 0 0}
\\
& & \times
Y_{t_1 t_2}^k (\hat{\vec{p}-\vec{p}\,'}, \hat{\vec{q}-\vec{q}\,'}),
\end{eqnarray*}
$$ \eqno{(A.3)} $$
where
$$F(a,b,c)=\sqrt{\frac{(2a+1)!}{(2b)!(2c)!}}.$$
Denote by $f_{a}(k_{a})$ the product of the propagator
of meson $a$ by the two $NNa$ couplings, including the form factor
function that is introduced at the vertices.
The momentum $k_{a}$ depends on
three angles: the angle between
${\vec p -\vec p\,'}$ and ${\vec q -\vec q\,'}$, whose cosine is $x_1$,
%$x_1=\hat{\vec p -\vec p\,'} \cdot \hat{\vec q -\vec q\,'}$,
the angle between $\vec p$ and $\vec p\,'$, the cosine
of which is $x_2$,
%$x_2=\hat p \cdot \hat p'$
and the angle
between $\vec q$ and $\vec q\,'$, the cosine of which is $x_3$.\\

To prepare the angular integrations we can start
by doing a decomposition in Legendre polynomials in
the angular variable
$x_1$:
$$ f_{\pi}(k_{\pi}) f_{\sigma}(k_{\sigma}) =\sum_{l_1}
\frac{2l_1+1}{2} g_{l_1} (|\vec p -\vec p\,'|,|\vec q - \vec q\,'|)
P_{l_1}(x_1) \eqno{(A.4)}$$

The two additional Legendre polynomial decompositions --- in
$x_2$ and $x_3$ --- involve the function
$g_{l_1} (|\vec p -\vec p\,'|,|\vec q - \vec q\,'|)$ of Eq.(A.4) as well as
powers of $|\vec p -\vec p\,'|$ and $|\vec q - \vec q\,'|$,
originated by contracting the function
$Y_{t_1 t_2}^k (\hat {\vec p-\vec p\,'}, \hat {\vec q-\vec q\,'})$ from Eq.
(A.3) with $P_{l_1}$
from Eq. (A.4).
The final  result of the 3 decompositions is
\begin{eqnarray*}
H_{l_1 l_2 l_3 \alpha_1 t_3 \alpha_2 t_4} (p,p',q,q') & = &
\int_{-1}^1 d x_1 \int_{-1}^1 d x_2 \int_{-1}^1 d x_3
P_{l_1}(x_1)  P_{l_2}(x_2)  P_{l_3}(x_3)
\\
& & \times
|\vec p -\vec p\,'|^{\alpha_1 -t_3}
|\vec q - \vec q\,'|^{\alpha_2-t_4}
f_\pi (k_\pi) f_\sigma (k_\sigma)
\, .
\end{eqnarray*}
$$ \eqno{(A.5)} $$
Using Eq. (A.1), each term of the $\sum_k$,
$ V_{\pi\sigma}^k$, gives for the orbital part of the matrix element,
\begin{eqnarray*}
\lefteqn{
\langle (l' \lambda') L' M' | V_{\pi\sigma}^k | (l \lambda) L M \rangle
=
\frac{ (4\pi)^2}{8}
\hat{k}^2 \hat{L} \quad C^{k L L'}_{\mu M M'}
(-)^{k+l+\lambda}
\sum_{l_1 l_2 l_3}
(-)^{l_1+ l_2+ l_3}
}
\\
& & \times
\hat{l}_1^2 \hat{l}_2^2 \hat{l}_3^2
\sum_{t_1 t_2 t_3 t_4} \hat{t}_1 \hat{t}_2 \hat{t}_3 \hat{t}_4
\quad
\xi^{l_1 l_2 l_3 t_3 t_4}_{t_1 t_2 k}
\left\{ \begin{array}{ccc}
t_2 & t_1 & k\\
t_3 & t_4 & l_1
\end{array}  \right\}
\left\{ \begin{array}{ccc}
l & l' & t_3 \\
\lambda & \lambda' & t_4 \\
L & L' & k
\end{array} \right\}
C^{t_1 l_1 t_3}_{0 0 0}
C^{t_2 l t_4}_{0 0 0}
\\
& & \times
\phi_{t_3 l_2 l l'}(p,p')
\phi_{t_4 l_3 \lambda \lambda'} (q,q')
\, ,
\end{eqnarray*}
$$ \eqno{(A.6)} $$
with
$$
\phi_{abcd}(k,k') =
\sum_{f_1+f_2 = a} F(a,f_1,f_2) k^{f_1} k'^{f_2} (-)^{f_2}
C^{b f_1 c}_{0 0 0} C^{b f_2 d}_{0 0 0}
\left\{ \begin{array}{ccc}
d & c & a \\
f_1 & f_2 & b
\end{array}  \right\}
$$
$$ \eqno{(A.7)} $$
and
\begin{eqnarray*}
\xi^{l_1 l_2 l_3 t_3 t_4}_{t_1 t_2 k} & = &
\sum_{r_1+r_2=1}
\sum_{s_1 +s_2=1}
C^{r_1 s_1 t_1}_{0 0 0}
C^{r_2 s_2 t_2}_{0 0 0}
F(1,r_1,r_2) F(1,s_1,s_2)
(-)^{r_2}
\\
& & \times
\left( \frac{1}{2} \right)^{r_2+s_2}
\left\{ \begin{array}{ccc}
r_1 & r_2 & 1 \\
s_1 & s_2 & 1 \\
t_1 & t_2 & k
\end{array}  \right\}
H_{l_1 l_2 l_3 \alpha_1 t_3 \alpha_2 t_4} (p,p',q,q')
\, .
\end{eqnarray*}
$$ \eqno{(A.8)} $$
($\alpha_1=r_1+s_1$; $\alpha_2=r_2+s_2$).

Applying the Wigner-Eckart theorem, we extract from Eq.(A.6)
the orbital reduced
matrix element, which ultimately we reconnect to the spin one, generating
the desired final result.
\newpage

\centerline{\bf Table I}

\vspace{1cm}

\begin{center}
\begin{tabular}{|l|r|r|} \hline
 & H.O. & MF I-III\\ \hline
$V_{\pi \sigma}$ & 0.392 & 0.457\\
$V_{\pi a}$ & --0.012 & --0.010\\
$V_{\pi\omega}$ & --0.301 & --0.337\\
$V_{\pi \rho}^C$ & 0.020 & 0.016\\ \hline
TOT & 0.099 & 0.126 \\ \hline
\end{tabular}
\end{center}

\vspace{0.5cm}

Matrix elements in MeV of the $\pi$-scalar and $\pi$-vector exchange
interactions which involve an intermediate nucleon-antinucleon pair. The
three-nucleon interactions are constructed from the Bonn boson exchange
model OBEPQ for the $NN$ interaction \cite{Bonntab5}. The matrix
elements are obtained with $S$-state oscillator and Malfliet-Tjon I-III
wavefunctions.

\vspace{0.5cm}

\centerline{\bf Table II}

\vspace{1cm}

\begin{center}
\begin{tabular}{|l|r|r|} \hline
 & PARIS & OBEPQ\\ \hline
$V_{\pi \sigma}$ & 0.551 & 0.836\\
$V_{\pi\omega}$ & --0.417 & --0.642\\ \hline
TOT & 0.134 & 0.194 \\ \hline
\end{tabular}
\end{center}

\vspace{0.5cm}

Matrix elements in MeV of the $\pi$-$\sigma$ and $\pi$-$\omega$ exchange
interactions which involve an intermediate nucleon-antinucleon pair. The
three-nucleon interactions are constructed from the Bonn boson exchange
model OBEPQ for the $NN$ interaction \cite{Bonntab5}. The matrix
elements are obtained with the $S$-state components of the Paris and
Bonn (OBEPQ) wavefunctions.

\vspace{1cm}
\noindent
{\bf Figure caption}
\vspace{0.5cm}

Fig. 1 \hspace{0.2cm} (a) $\pi$-short range exchange three-nucleon
interaction that involves an intermediate nucleon-antinucleon pair,
(b) $\pi$-short range exchange three-nucleon
interaction that involves excitation of an intermediate $N(1440)$
resonance. The wavy lines symbolize scalar and vector meson
exchange.\\


\newpage

\begin{thebibliography}{999}

\bibitem{BG} G.\ E.\ Brown and A.\ M.\ Green, Nucl. Phys. {\bf A137}
1, (1969).
\bibitem{reviews} B.\ F.\ Gibson and B.\ H.\ J.\ McKellar, Few-Body
Systems {\bf 3}
(1988) 143; S.\ A.\ Coon and M.\ T.\ Pe\~{n}a, Few-Body Systems, {\bf
Suppl. 6}, 242 (1992).
        \bibitem{TM1} S.\ A.\ Coon, M.\ D.\ Scadron, P.\ C.\ McNamee, B.\ R.\
Barrett, D.\ W.\ E.\ Blatt and B.\ H.\ J.\ McKellar, Nucl. Phys. {\bf
A317}, 242 (1979); S.\ A. Coon and W.\ Gl\"{o}ckle, Phys. Rev. {\bf C23}
1790 (1981).
        \bibitem{TM2} R.\ G.\ Ellis, S.\ A.\ Coon, and B.\ H.\ J.\ McKellar,
Nucl. Phys. {\bf A438}, 631 (1985); S.\ A.\ Coon and M.\ T.\ Pe\~{n}a,
Phys. Rev. {\bf C48} 2559 (1993).
       \bibitem{Chen} C.\ R.\ Chen, G.\ L.\ Payne, J.\ L.\ Friar, and
B.\ F.\ Gibson, Phys. Rev. {\bf C33}, 1740 (1986); T.\ Sasakawa and S.\
Ishikawa, Few-Body Systems {\bf1}, 3 (1986); J.\ Carlson, V.\ R.\
Pandharipande, and R.\ B.\ Wiringa Nucl. Phys. {\bf A401}, 59 (1983).
        \bibitem{overbind} S.\ A. Coon, J.\ Zabolitzky, and D.\ W.\ E.\
Blatt, Z. Physik A {\bf 281}, 137 (1977); R.\ B.\ Wiringa,
Phys. Rev. {\bf C43}, 1585 (1991); H.\ Kamada and W. Gl\"{o}ckle, Nucl.
Phys. {\bf A560}, 541 (1993).
        \bibitem{phTNI} R. Schiavilla, V.R. Pandharipande and R.B.
Wiringa, Nucl. Phys. {\bf A449}, 219 (1986).
        \bibitem{Stadler} A.\ Stadler, Bull. Am. Phys.
Soc. {\bf 38}, 1013 (1993);  A. Stadler, J.\ Adam Jr.,
J.\ Henning, and P.\ U.\ Sauer,  ``$\pi$ and $\rho$-exchange
Three-Nucleon Forces in the Three- Nucleon Bound State", Proc.
14th European Conference on Few-Body Problems in Physics, Amsterdam,
23-27 August, 1993, Conference Handbook ISBN 90 5294 080 0,
p. 192; A. Stadler, J.\ Adam Jr.,
J.\ Henning, and P.\ U.\ Sauer,  ``$\pi$ and $\rho$-exchange
Three-Nucleon Forces in the Three- Nucleon Bound State", submitted
to Phys. Rev C.
       \bibitem{GT}S.\ A.\ Coon and M.\ D.\ Scadron,   Phys. Rev. {\bf
C23}, 1150 (1981); ibid {\bf C42}, 2256 (1990); $\pi$-N Newsletter, {\bf
3}, 90 (1991).
        \bibitem{Holinde} J. Haidenbauer, K.\ Holinde, and A.\ W.\
Thomas, Phys. Rev. {\bf C45}, 952 (1992); G. Janssen, K.\ Holinde, and
J.\ Speth, Phys. Rev. Lett. {\bf 73}, 1332 (1994).
        \bibitem{Keister} B.D. Keister and R.B. Wiringa, Phys. Lett.
{\bf 173}, 5 (1986).


        \bibitem{Blunden} P.G. Blunden and D.O. Riska, Nucl. Phys.
{\bf A536}, 697 (1992).

\bibitem{Robilotta}  M.\ R.\ Robilotta, Few-Body Systems,
      Suppl.{\bf 2} 2, 35 (1987).
\bibitem{FM} J.\ Fujita and H.\ Miyazawa, Prog. Theor. Phys. {\bf 17},
360 (1957).
        \bibitem{Argonne} O.\ Benhar, V.\ R.\ Pandharipande and S.\ Peiper,
Rev. Mod. Phys. {\bf 65}, 817 (1993); R.\ B.\ Wiringa, ibid. 231. See also
[5c] and \cite{phTNI}.

        \bibitem{Malfliet} J.\ L.\ Friar, B.\ F.\ Gibson,  and G.\ L.\
Payne, Z. Phys. {\bf A301}, 309 (1981); R. Malfliet and J. Tjon, Nucl.
Phys. {\bf A127}, 161 (1969).

        \bibitem{Alfred} A.\ Stadler, W. Gl\"{o}ckle and P.\ U.\ Sauer,
 Phys. Rev. C {\bf 44}, 2319 (1991).

        \bibitem{Bonntab5} R.\ Machleidt, K.\ Holinde and Ch.\ Elster,
Phys. Rept. {\bf 149}, 1 (1987)
        \bibitem{Nijmegen} V.\ G.\ J.\ Stoks, R.\ A.\ M.\ Klomp, C.\ P.\
F. Terheggen, and J.\ J.\ de Swart, Phys. Rev. C {\bf 49}, 2950 (1994);
M.\ M.\ Nagels, T.\ A.\ Rijken, and  J.\ J.\ de Swart, Phys. Rev. D {\bf
17}, 768 (1978).

        \bibitem{Paris} M.\ Lacombe et al., Phys. Rev. {\bf C21}, 861
(1980).
        \bibitem{Charap} J.\ M.\ Charap and M.\ J.\ Tausner, Nuovo Cim. {\bf
18}, 316  (1960).
        \bibitem{Chemtob} M.\ Chemtob, J.\ W.\ Durso and D.\ O.\ Riska,
Nucl. Phys. {\bf B38}, 141 (1972).

        \bibitem{pppi} T.-S.\ H.\ Lee and D.\ O.\ Riska, Phys. Rev.
Lett. {\bf 70}, 2237 (1993); C.\ J.\ Horowitz, H.\ O.\ Meyer, and D.\
K.\ Griegel, Phys. Rev. C {\bf 49}, 1337 (1994).

        \bibitem{Sasrho} T.\ Sasakawa, S.\ Ishikawa, Y.\ Wu, and T-Y.
Saito, Phys. Rev. Lett. {\bf 68}, 3503 (1992).

        \bibitem{Pena}S.\ A.\ Coon,  M.\ T.\ Pe\~{n}a, R.\ G.\  Ellis,
 Phys. Rev.  {\bf C30}, 1366 (1984).

	\bibitem{Friar} J.\ L.\ Friar, Phys. Rev. C {\bf 22}, 796
(1980).
	\bibitem{CoonFriar} S.\ A.\ Coon and  J.\ L.\ Friar, Phys. Rev. C {\bf
34}, 1060 (1986); J.\ L.\ Friar and S.\ A.\ Coon, Phys. Rev. C {\bf 49},
1272 (1994).
	\bibitem{Adam}  J.\ Adam, Jr., H.\ G\"{o}ller, and A.\ Arenh\"{o}vel,
Phys. Rev. C {\bf 48}, 370 (1993); B. Desplanques and A.\ Amghar, Z.
Phys. A {\bf 344}, 191, (1992).

        \bibitem{Gerjuoy} E.\ Gerjuoy and J.\ Schwinger, Phys. Rev. {\bf
61}, 133 (1942).


\end{thebibliography}
\end{document}